\documentclass[12pt,preprint]{aastex}

\newcommand{\qaa}{J131046+0006}
\newcommand{\qab}{J131055+0008}
\newcommand{\qba}{J135457-0034}
\newcommand{\qbb}{J135504-0030}
\newcommand{\qca}{Q 0107-0235}
\newcommand{\qcb}{PB 6291}
\newcommand{\qda}{J011441-3139}
\newcommand{\qdb}{J011446-3141}
\newcommand{\qpa}{QP1310+0007}
\newcommand{\qpb}{QP1355-0032}
\newcommand{\qpc}{QP0110-0219}
\newcommand{\qpd}{QP0114-3140}

\shorttitle{Searching High Redshift Large-Scale Structures}
\shortauthors{Boris et al.}

\begin{document}

\title{Searching High Redshift Large-Scale Structures:\\
    Photometry of Four Fields Around Quasar Pairs at $z \sim 1$}

\author{N. V. Boris\altaffilmark{1}, L. Sodr\'e Jr.\altaffilmark{1},
  E. S. Cypriano\altaffilmark{2}, W. A. Santos\altaffilmark{1},
C. Mendes de Oliveira\altaffilmark{1} \\ and M. West\altaffilmark{3}}

\altaffiltext{1}{Departamento de Astronomia, Instituto de Astronomia, Geof\1sica e Ci\^encias Atmosf\'ericas, Universidade de S\~ao Paulo,
Rua do Mat\~ao 1226, Cidade Universit\'aria 05508-900, S\~ao Paulo, SP, Brazil; natalia@astro.iag.usp.br}
\altaffiltext{2}{Department of Physics and Astronomy, University College London, Gower Street, London, WC1E 6BT,
United Kingdom}
\altaffiltext{3}{Gemini Observatory/AURA, Southern Operations Center,
Casilla 603, La Serena, Chile}

\begin{abstract}

We have studied the photometric properties of four fields around the
high-redshift  quasar pairs \qpa, \qpb, \qpc, and \qpd ~at z $\sim$
1 with the aim of identifying large-scale structures- galaxy
clusters or groups- around them. This sample was observed with GMOS
in Gemini North and South telescopes in  the $g'$, $r'$, $i'$, and
$z'$ bands, and our photometry is complete to a limiting magnitude
of $i' \sim 24$ mag (corresponding to $\sim M_{i'}^* + 2$ at the
redshift of the pairs). Our analysis reveals that \qpc~
shows very strong and \qpa~ and \qpb~ show some evidence for the presence 
of rich galaxy clusters in direct vicinity of the pairs.
On the other hand, \qpd~ could be an isolated pair in a poor environment.
This work suggest that $z \sim 1$
quasar pairs are excellent tracers of high density environments and this
same technique may be useful to find clusters at higher redshifts.
\end{abstract}

\keywords{galaxies: clusters: general --- galaxies: distances and redshifts ---
galaxies: high-redshift --- large-scale structure of universe ---
quasars: general --- X-ray: galaxies: clusters}

\section{Introduction}

The study of galaxy populations in high-redshift large-scale structures
can give us important clues about the star formation history and
galaxy formation process in such
environments \citep{wf91,bekki98,kodama98,vandok05,mei06}.
However, the detection of distant  galaxy clusters is not trivial.

At high redshifts, the use of techniques like the red-cluster sequence
\citep{RCS}, is less efficient. The reason
is the proportional decrease of the number of red galaxies in clusters
for increasing redshift \citep[e.g.][]{bo84}. Therefore, in order to detect
high-z clusters through optical imaging, one requires at least photometric
redshift information, which implies imaging in four or more bands.
Additionally, in order to avoid the high observational cost of observing large
areas of the sky, one can use several indicators of the presence of high-z
clusters to select the fields to be observed.

Several of these tracers, like extended X-ray emission
\citep[e.g.,][]{romer01}, the Sunyaev-Zeldovich
decrement or bright radio-emitting galaxies have been largely used
to trace clustering of galaxies.
These techniques have  strengths and drawbacks.
The first two, for example,
depend on the presence of a hot intra-cluster medium, which may bias samples
against recently forming clusters.
In this paper, we will consider another possible tracer of the presence of clusters:
physically close pairs of quasars.

Quasars are relatively rare astronomical objects  and hence, if they are
distributed following galaxies, the presence of two or more such objects in
a relatively small volume should be a good indicator of a rich environment.
Actually, in structure formation scenarios with bias between barionic and dark
matter distribution \citep{kaiser84} it is expected that high redshift objects
form in  large high--redshift density fluctuations and, therefore, such correlation
between quasar concentration and clusters is somewhat expected, unless, 
for some reason, quasars avoid clusters.
However, most observational evidence shows that high redshift quasars do tend to follow
the overall large scale structures.

Whether quasars inhabit or not high density regions in low redshifts is a
subject of dispute.
\citet{col02}, for example, claim that at $0.1 \le z \le 0.25$,
quasars (both radio--loud and radio--quiet) tend to reside in low density
regions. On the other hand \citet{mullis}, using a sample of X-ray selected
quasars, conclude that those objects trace closely the underlying mass
distribution. \citet{ilona02} also points out that $0.2<z <0.3$
quasars follow the large-scale structure traced by galaxy clusters, but
they also note the complete absence of radio--quiet QSO's at the very
center of galaxy clusters.

At higher redshift, however, most observational results suggest that
quasars prefer groups or clusters \citep{HG98,wold00, wold01}.
One very convincing example is the structure found by \citet{haines01}
at $z = 1.226$ around a radio-quiet quasar belonging to a large quasar
structure \citep{cc91,cc94}.
The same behaviour appears to be followed by radio-loud quasars.
A good example is the work by \citet{sg02}, who found a highly significant excess of
galaxies around radio-loud quasars at $1.0 < z < 1.6$. \citet{tanaka01}
also points
in the same direction by reporting an overdensity of galaxies around a quasar
concentration at $z\sim1.1$. An exception is the work by
\citet{coil07} who, through an analysis of the clustering  of quasars and 
galaxies at $0.7 < z < 1.4$, concluded that quasars and blue galaxies are 
found in the same environment, which differs from that occupied by the 
red galaxy population. Regarding specifically quasar pairs,
\citet{zs01} found an statistically significant excess of high-redshift
quasar pairs with
separations between 1 and 5 Mpc in projected distance. This suggests that such
quasar pairs belong to sizable physical structures (precursors of today's
clusters and
superclusters of galaxies) and therefore, they can be used as tracers of
high-redshift large-scale structures. Going to even larger redshifts,
\citet{dj03}  found that a quasar pair at $z = 4.96$ is associated with a
large-scale  structure. Thus, an interesting form to search for high-redshift
clusters and other large-scale structures is examining the environment
inhabited by quasar pairs.

In this work, we describe a multi-color photometric study of the field around
four quasar  pairs at $z \sim 1$, using the instrument GMOS in both Gemini
North and South telescopes.

One of the pairs in our sample, QP0110-0219, has been previously studied
by \citet{surdej86}, who found hints
of the presence of a cluster around it.
The new data we present here allow us to confirm 
this claim. There are no studies in the literature for the
other three quasar pairs.

The outline of this paper is the following: in Section 2 we describe the sample
and the data reduction procedures. The galaxy photometry is discussed in
Section 3. Section 4 outlines our approach to obtain photometric redshifts
and presents then application to our galaxy sample. The environments of the
quasar pairs are discussed in section 5. Finally, in Section 6 we summarize our
results. Throughout this paper, we adopt a $\Lambda CDM$ concordance cosmology
with  $\Omega_m = 0.3$ and $\Omega_{\Lambda} = 0.7$, and we use the value $h =
0.7$ in the Hubble constant, $H_0 = 100 ~h$ km s$^{-1}$ Mpc$^{-1}$.

\section{Observations and data reduction}

\subsection{Sample selection}

In this paper, we study a sample of four fields around quasar pairs at $z \sim 1$.
We selected the pairs from \citet{vv01} quasar catalog considering
redshift differences smaller than 0.01 and projected angular separations
smaller than 300 arcsec. We did not consider pairs with angular separations
 smaller than 15 arcsecs to avoid including gravitational lens. With
these parameters, we found 84 quasar pairs. Five of them had redshifts
between 0.9 and 1.0, and  four were observed with Gemini
telescopes. The main sample characteristics are shown in Table \ref{car}. It
includes: the quasar names, their coordinates, redshifts, angular
separation, and the name adopted for the pairs in this paper.

We have checked the spectra of the quasars in our sample
to certify that we indeed did not pick any cases of gravitationally
lensed images of one only quasar.
The parity of \qpc ~is discussed by \citet{surdej86}. Considering the
redshift difference and spectral characteristics, they conclude that the
quasars \qca ~and \qcb ~are different objects. For
the other pairs, quasar spectra are available in the 2dF QSO Redshift Survey
\citep{croom04}\footnote{http://www.2dfquasar.org/Spec\underline{~}Cat/2qzsearch2.html}.
Our visual examination of the spectra indicates that also in this case the
differences in redshifts and spectral characteristics suggest that they are
indeed different objects and not lensed images of the same quasar. Moreover, in \qpa, \qpb,
and \qpc ~one of the quasars is radio-loud and the other is radio-quiet.
Consequently, we are confident that none of the pairs in our sample are
produced by gravitational lensing.

\subsection{Imaging and data reduction}

The four fields in  Table \ref{car} were observed with GMOS N and S
mounted on Gemini telescopes.
The imaging was done in four filters of
the SDSS system \citep{fuku96}: $g'$, $r'$, $i'$, and $z'$.
The log of observations is presented in Table \ref{obs2}, which shows the
telescope used, the exposure time, and the Gemini program identification number. All observations were performed in photometric conditions.
The typical FWHM for point sources was $\sim$ 0.7 arcsec in all images.

Data reduction was performed using the
Gemini IRAF \footnote{IRAF is distributed by the National Optical Astronomy
Observatories, which are operated by the Association of Universities for
Research in Astronomy, Inc., under cooperative agreement with the National
Science Foundation.} package.
The images were bias corrected, flat fielded, and fringe corrected in the
standard way. After that, they were combined and cleaned of cosmic ray events
and bad pixels producing, then, the final images, appropriate for science
analysis.

\section{Photometry and object detection \label{phot-detect}}

We have used the IRAF package {\it daophot}
to calculate the photometric zero-point for each band in each field in
the AB SDSS photometric system. The calibration was made using
stars from the Landolt catalog \citep{lan} also calibrated in
the AB SDSS system. Using the dispersion in
the magnitudes of the stars, we have estimated the accuracy of the magnitude
zero-point as 0.01 in $g'$ and $r'$ bands, 0.02 in $i'$ band, and
0.03 in $z'$ band.

We have used SExtractor \citep{sex} to detect objects over the final image frames.
First, we ran the program on the images of each photometric band and
selected the image that showed the highest  number of detected objects.
Second, using such image as reference, we ran the program again in "dual
image mode". We used a top-hat filter and detected objects above
1.5 $\sigma$, which corresponded to median isophotal levels of 27.1, 26.4,
26.4, and 25.4 mag arcsec$^{-2}$ in $g'$, $r'$, $i'$, and $z'$,
respectively. In order to run the program in "dual image mode", it was necessary to
align the images. Thus, because of rotations and shifts, parts of the
images near the borders were lost. 

Positions and magnitudes (total and aperture) were obtained for all objects
present in all bands for each pair. We adopted 3 arcsec aperture magnitudes,
$m_{ap}$, to measure colors. Aperture magnitudes were also obtined to compare
our data with others in the literature. For the total
magnitude of an object, we have addopted a color corrected isophotal
magnitude. For example, if the objects were detected
in the g' band,
then $g' = g_{iso}$, $r'= g_{iso} - (g_{ap}-r_{ap})$, etc.,
where $m_{iso}$ is the isophotal magnitude given by SExtractor. After
measuring the magnitudes they were corrected for Galaxy extinction, with the
absorption coefficients $A_{\lambda}$ obtained from
\citet{schlegel98} using NED and interpolated to GMOS bands.

\begin{figure}[h!]
\plotone{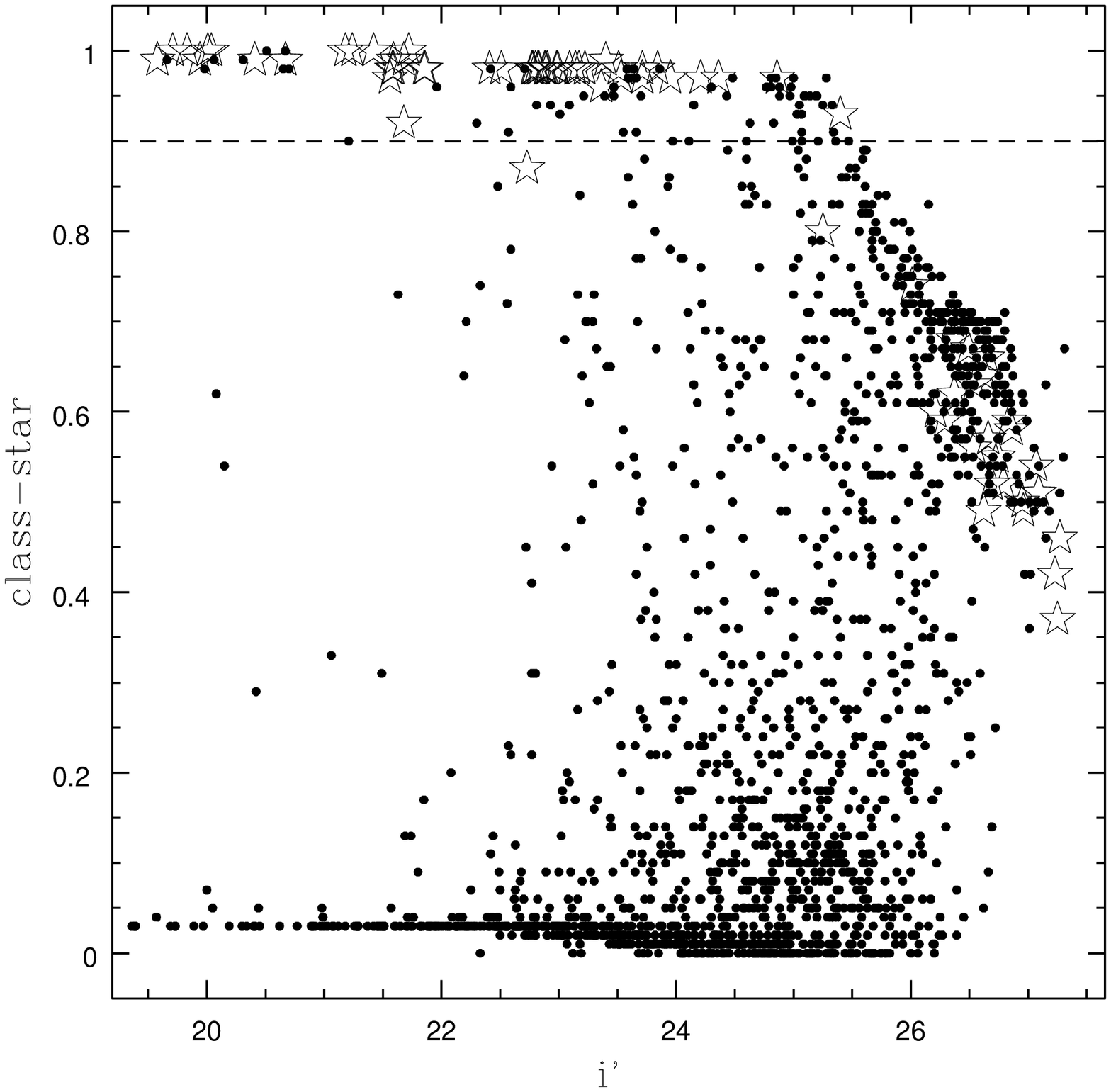} \caption{\label{cs}{ 
Star/Galaxy separation for the pair \qpc.
Stars represent objects with FWHM $\le$ seeing and filled circles represent
objects with FWHM $>$ seeing. We have considered as galaxies all objects
with class-star $<$ 0.9 (dashed line).}}
\end{figure}

We have used the class-star parameter of
SExtractor, which ranges from 0 (galaxies) to 1 (stars),
to separate stars from galaxies. 
Figure  \ref{cs} shows this parameter versus the $i'$ magnitude for the 
pair \qpc. The star symbol in the plot represents all objects with FWHM 
$\le$ seeing. If we consider all objects with class-star $<$ 0.8 as galaxies, 
a threshold often adopted in the literature \citep[e.g.][]{kodama04,kaputi06}, 
we find that 2 \% of the objects with FWHM $\le$ seeing
have class-star $<$ 0.8 and that 7 \% of the objects with
FWHM $>$ seeing have class-star $\ge$ 0.8. On the other hand,
if we adopt a threshold of 0.9, we have a similar contamination ($\sim$2 \%)
of the galaxy sample and only 4 \% of the objects with
FWHM $>$ seeing have now class-star $\ge$ 0.9. The results for the other 
fields are similar. We then decided to adopt the class-star value of 0.9 
to separate stars from galaxies. The same criteria was adopted by 
\citet{capak04} to determine number counts in the HHDFN.

In order to estimate the completeness magnitude of the observations, we have ploted the
logarithmic number of detected objects as a function of the total magnitude
in the band used for detection. From visual inspection of the
turnover magnitudes, we estimated that the observations are complete down to
$i'$ = 24 for \qpa ~and \qpc, $g'$ = 25 for \qpb, and $g'$ = 24.5 for \qpd.

It is interesting to know how the magnitudes above compare with those of
a $M^*$-galaxy at the redshift of the pairs. We have estimated the value
$M^*$ in two ways, as follows. 
\citet{ellis04} obtained
$K^* \sim 18$ for clusters of galaxies with redshifts between 0.8 and 1.0.
Considering the value for $(I-K) \sim 2.9$ obtained by \citet{stanford02}
for the cluster 3C 184 ($z = 0.996$), we have $I^* \sim 20.9$.
On the other hand, using spectrophotometric synthesis models, \citet{fuku95}
obtained $(i'-I_c) \sim 0.7$ for galaxies at $z = 0.8$. Then,
we obtain $i'^* \sim 21.6$.
In the second case, we may consider the Coma cluster as representative of a
$z = 0$ cluster. \citet{mobasher03} studied its luminosity  function and
found $M^*_R \sim -21.79 + 5 ~log ~h_{65}$. For galaxies at $z = 0$,
\citet{fuku95}
obtained $(r'-R_c) \sim 0.22$ and $(r'-i') \sim 0.30$, therefore
$M^*_{i'} \sim -21.71$ for the cosmology adopted here. We have calculated
$i'^*$ with
\begin{equation}
m = M(z=0) + 5 \log d_L[Mpc] + 25 + k(z) + e(z)
\end{equation}
where $d_L$ is the luminosity distance, $k(z)$ is the k-correction
and $e(z)$ is the evolution correction. In the cosmology adopted here,
$d_L \sim 6000 Mpc$ at $z \sim 1$. Using $k(z)$ and $e(z)$ values
published by \citet{fuku95} and \citet{po97} respectively ($k_{i'}
\sim 0.9$ and $e_{i'} \sim -1.3$) and the value obtained to
$M^*_{i'}$ at $z = 0$, we obtain $i'^* \sim 21.8$. Considering the
uncertainties of the approaches, the agreement of the two values is
very good. We have then adopted the mean value $i'^* \sim 21.7$.
This value is similar to that obtained by \citet{bla06} ($i^*_{775}
= 22.0 \pm 0.1$ AB) for early-type galaxies at $z = 0.83$.  This
result shows that the completeness magnitude of our fields
corresponds to  $\sim M^* + 2$ at the redshift of the pairs.

\subsection{Comparison with HHDFN and ACS-GOODS photometry \label{comp}}

Our approach to compute photometric redshifts (\S \ref{zphot}) makes use of a
training set with galaxies of known redshifts measured in the same photometric
bands. Consequently, photometric redshifts are very sensitive to small
zero-point changes. In order to examine this point, we have made a comparison
between our photometry with those available for the  HHDFN (Hawaii Hubble Deep
Field North) region \citep{capak04} and for the ACS-GOODS (Advanced Camera
for Surveys - Great Observatories Origins Deep Survey) region
\citep{cowie04}.

Although one region is contained in the other, the
photometric and spectroscopic data  available for them are different.
A comparison with HHDFN is useful because it contains
ACS-GOODS and has a photometric completeness similar to those of our fields.
ACS-GOODS, on the other hand, has hundreds of mesured spectroscopic redshifts and will be
adopted as the training set for our photometric redshift method.
Since the photometry of these fields (in UBVRIz$^\prime$) is diferent
of our photometry (in $g'r'i'z'$), 
they were interpolated to the GMOS bands adopted here.

First we have considered the HHDFN region
which has photometry complete down to $R = 24.5$.
We compared the photometric distribution of galaxies in our fields
within the magnitude completeness limit of each band with the corresponding
distributions using the interpolated magnitudes. We estimated the shift that
is required in the zero-point of our photometric bands so that the median
of the magnitude distribution (for galaxies brighther
than the completeness limit)
of  HHDFN and ours match each other.
We have used an iterative algorithm,
adding to our magnitudes a shift obtained in each step
until convergence. 
The median of the absolute value of the shifts is 0.07.

Many are the possible sources of
these shifts. A possibility is the
magnitude interpolation required in this approach.
Another one is cosmic variance, since
the galaxy catalogues considered here are not large enough. Indeed,
a photometric study with SDSS data made by \citet{fuku04} shows that,
besides Galaxy extinction, the principal cause for variations in number
counts is the large-scale clustering of galaxies. This dispersion increases for
smaller areas, being greater than 0.2 magnitudes for areas smaller than 0.01
$deg^2$, as is our case.

After appling the zero-point shifts in our magnitudes, we compared our
galaxy number counts with those from the ACS-GOODS
data in 0.5 mag intervals for objects brighter than the apperture
magnitude $z' = 22$. It can be verified that, for all fields, we
obtain a  good match between our number counts and those
of  ACS-GOODS. 

\section{Photometric redshift analysis} \label{zphot}

Determining the redshifts of the galaxies in our fields allows us to
separate the galaxies belonging to a possible cluster or group at the
redshift of the pair from the foreground and/or background galaxies.
Here we adopt photometric redshifts for this task.

Photometric redshift estimation is often done by comparing the magnitudes
of an object with the magnitudes of templates obtained with spectrophotometric
evolution models, as is the case of Zpeg \citep{zpeg} and HyperZ \citep{bol}.
Here we adopt another approach: instead of a galaxy model, we use real data-
magnitudes and spectroscopic redshifts- obtained in galaxy surveys.
We compare the magnitudes of our galaxies with magnitudes in the same bands
of real galaxies with known spectroscopic redshift to parametrize a local
empirical relationship between magnitudes or colors with redshift. This is
done using a locally weighted regression algorithm (LWR) developed
by our group (Santos et al. 2007, in preparation). The same type
of data-driven approach is adopted in the ANNz photometric redshift
package  \citep{annz}, which applies instead artificial
neural networks for this task.

\subsection{Method}
LWR is an algorithm designed to provide a continuous non-linear mapping
between sets of variables (e.g., Atkeson, Moore \& Schall 1997).
Our LWR method is discussed in detail and compared with other methods in
 Santos et al. (2007, in prep.).  Here we only
outline its main characteristics. The method works with magnitudes or colors
(and even with other galaxy properties, like diameter or type) but here we
use colors.

The method works with two data sets: the
training set (having known spectroscopic redshifts) and the test set
(for which we want to calculate the redshift). Obviously, both sets
must have the magnitudes and/or colors measured in the same bands and
in the same photometric system.

LWR establishes a linear relationship between colors
and redshifts that is local because the redshift estimation in a given point
in color space weights more heavily the data points in the neighborhood of
this point than those more distant.
The training set contains colors and spectroscopic redshift for all objects.
From these values we  build a redshift estimator which will be applied to
our galaxies, in the test set. We assume that the local relation between colors
and redshifts is linear:
\begin{equation}
\label{lin}
z({\bf x}) = a_0 + {\bf a}^T.{\bf x}= a_0 + \Sigma_{i=1}^{n} a_i x_i
\end{equation}
where ${\bf x}$ is a vector containing the $n$ colors of a given object,
$z({\bf x})$ is the redshift and $T$ stands for transpose matrix.
For each object in the test set, with colors in point ${\bf x}$,
we determine the values of coefficients $a_0 ... a_n$ and then the redshift by
minimizing the weighted $\chi^2$ function with the $N$ objects of the training
set:
\begin{equation}
\chi^2 = \Sigma_{j=1}^{N} \omega_j^2 \left(y_j-a_0-{\bf a}^T . {\bf x}_j\right)^2
\end{equation}
where $\omega_j$ is the weight associated to the $j-$th data point.
The locality of the fitting is assured by adopting a
weight function which decreases as the euclidean distance $d({\bf x},{\bf  x}_j)$
between points ${\bf x}$ and ${\bf  x}_j$ increases:
\begin{equation}
\label{k}
\omega_j = \exp\left(\frac{-d^2({\bf x},{\bf x}_j)}{2K^2}\right)
\end{equation}
The $K$ parameter is a kernel-width that determines the ``effective volume''
around point ${\bf x}$: only points within this sphere effectively affect the
values of the parameters and the redshift estimate. This parameter was determined in
this work by dividing randomly the training set objects in 2/3 for
training (TS) and 1/3 for validation (VS). 
For galaxies in VS we computed $z_{phot}$ with
eq. \ref{lin} using galaxies in TS to obtain the coefficients. 
This procedure was then repeated one hundred times.
For each realization of TS and VS, we compute the rms 
square deviation between $z_{phot}$
and $z_{spec}$ for the objects in VS, $\sigma_z$, and choose as 
optimal $K$ the value for
which $\sigma_z$ is minimum.

It is worth mentioning that redshift estimates with the LWR method are heavily
dependent on the training set adopted (besides, of course, the set of colors
available). In particular, the redshift accuracy increases with the size of the
training set and depends strongly on the homogeneity of the photometric
calibration of the training and test sets.

\subsection{Application to our sample} \label{aplic}

We adopt in this work the (interpolated) photometric data and spectroscopic
redshifts of the ACS-GOODS region as our training set.
Since the method allows  using colors or magnitudes for photometric
redshift estimation, we have used colors (but none of the results reported
in the next section depend of this choice). Using
magnitudes we would have to limit our sample at $z' = 22$ (the ACS-GOODS
spectroscopic completeness), but our sample photometry goes deeper; it is
complete at least down to $i' = 24$.
When we use colors, such limit is not necessary and we are able to
estimate  photometric redshifts for fainter objects, even without
spectroscopic data for $z' > 22$.

\begin{figure}[h!]
\plottwo{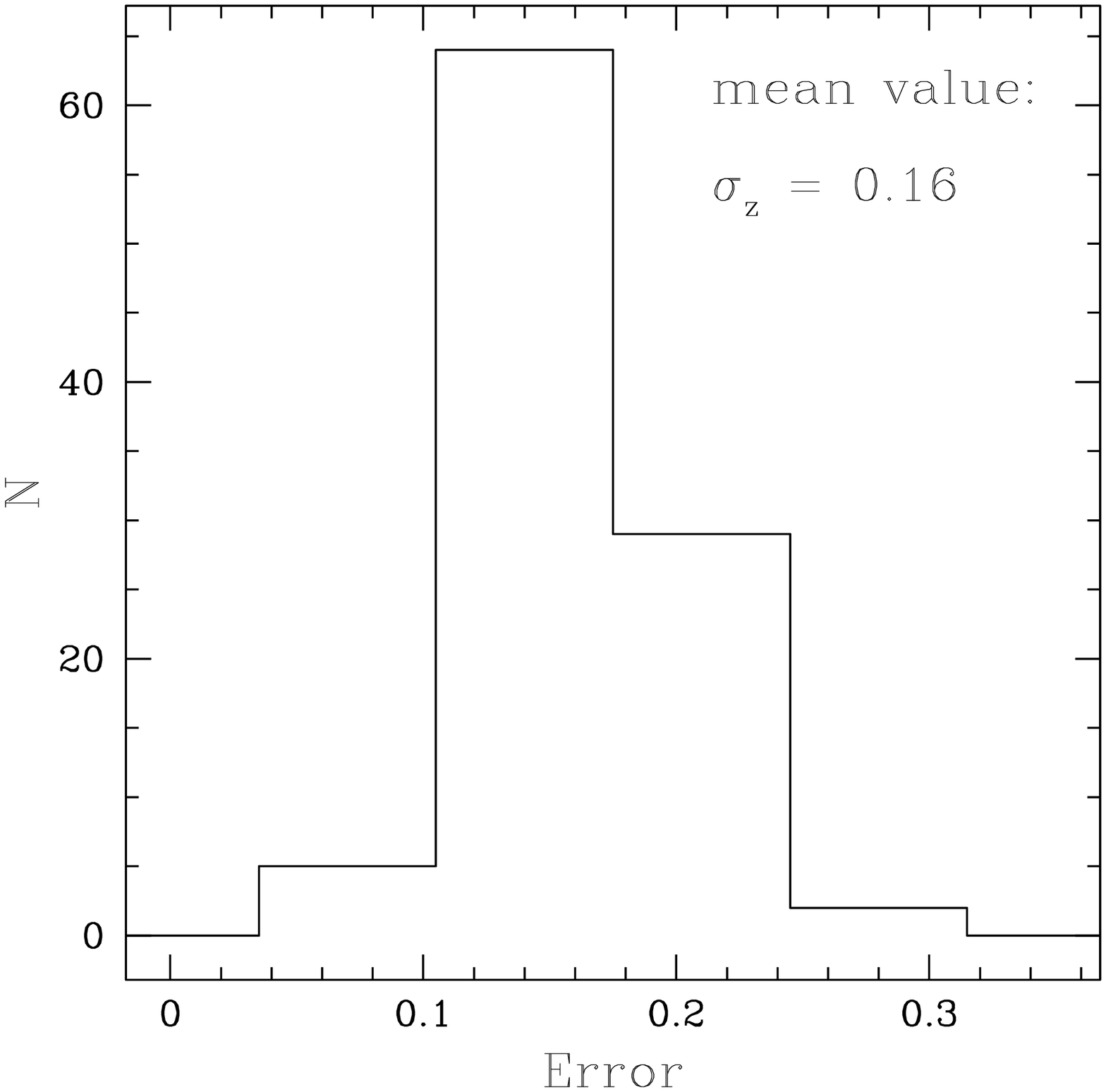}{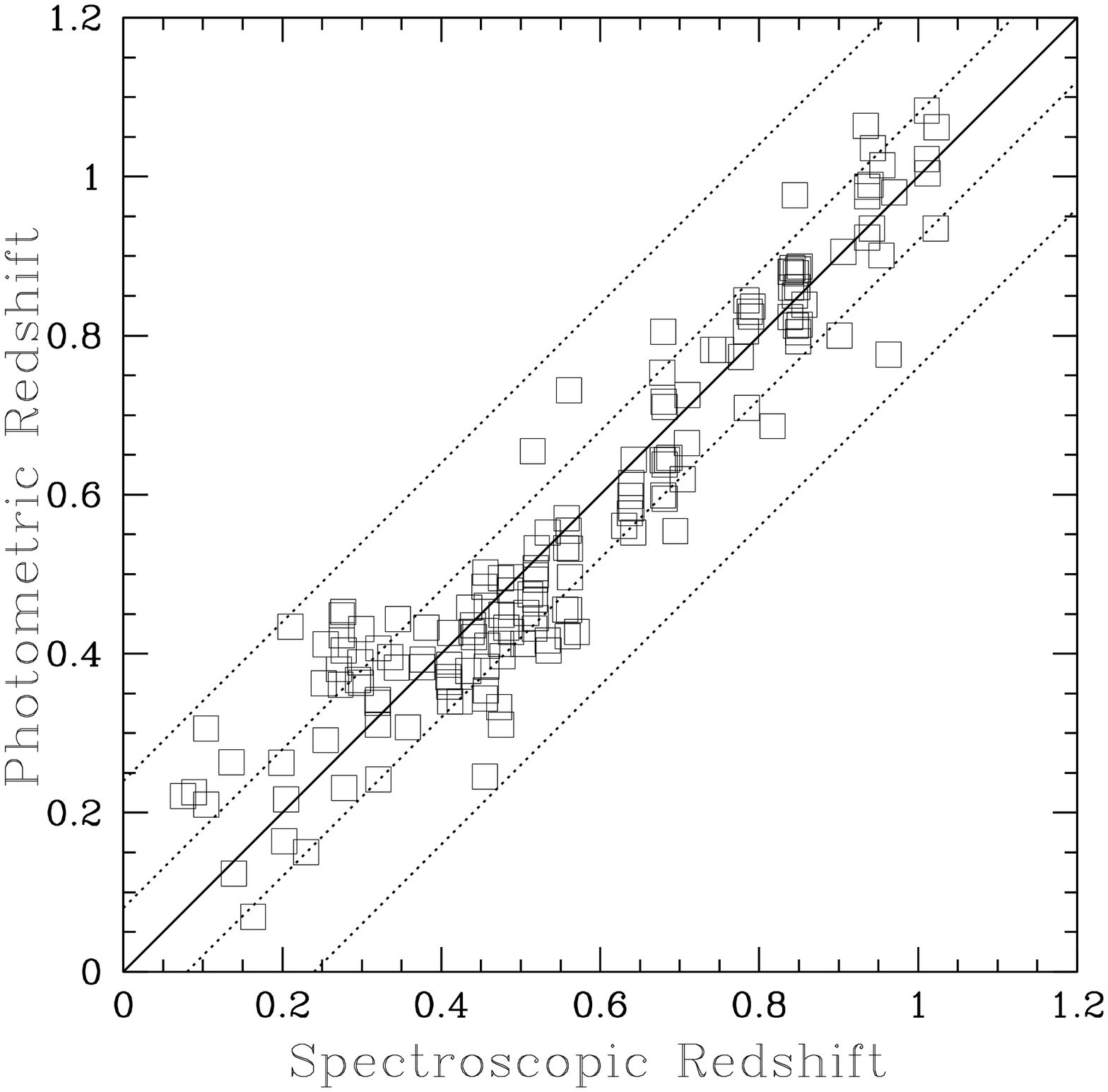}
\caption{\label{zph}{Left: photometric redshift error for 100
simulations. Right: comparison between photometric and spectroscopic
redshifts for the 1/3 of galaxies from ACS-GOODS used for validation,
corresponding to the simulation with photometric redshift error equal to
$\sigma_z=0.16$; the continuous line is the equality line and the
dashed lines correspond to 1 and 3 $\sigma_z$.}}
\end{figure}

The value $K=0.33$ was determined by the procedure described
in the previous section as the median value of 100 simulations.
The histogram in Figure \ref{zph} shows the redshift error distribution
for all simulations.  In what follows we consider the mean value,
$\sigma_z = 0.16$, as the redshift error for the training
set ACS-GOODS. Figure \ref{zph} also shows the comparison between
$z_{phot}$ and $z_{spec}$ for the 1/3 of galaxies from ACS-GOODS
used for validation corresponding to the simulation with this mean value.

Having obtained photometric redshifts for all fields containing quasar pairs,
we may start looking for structures around the redshift of the pairs, that is,
objects with $z_{phot} = z_{pair} \pm \Delta z$. We made experiments with
$\Delta z$ equal to 0.1, 0.16 and 0.2, obtaining similar results. Therefore we
only present here results  with $\Delta z = \sigma_z = 0.16$. For comparison,
\citet{toft03},  in a photometric study of the galaxy cluster MG2016+112 at $z
= 1$, adopted $\Delta z = 0.25$. Note that the error $\sigma_z$ should depend
of the photometric errors, the number of colors, and the size of the training set.

\section{Results}

\begin{figure}[h!]
\plotone{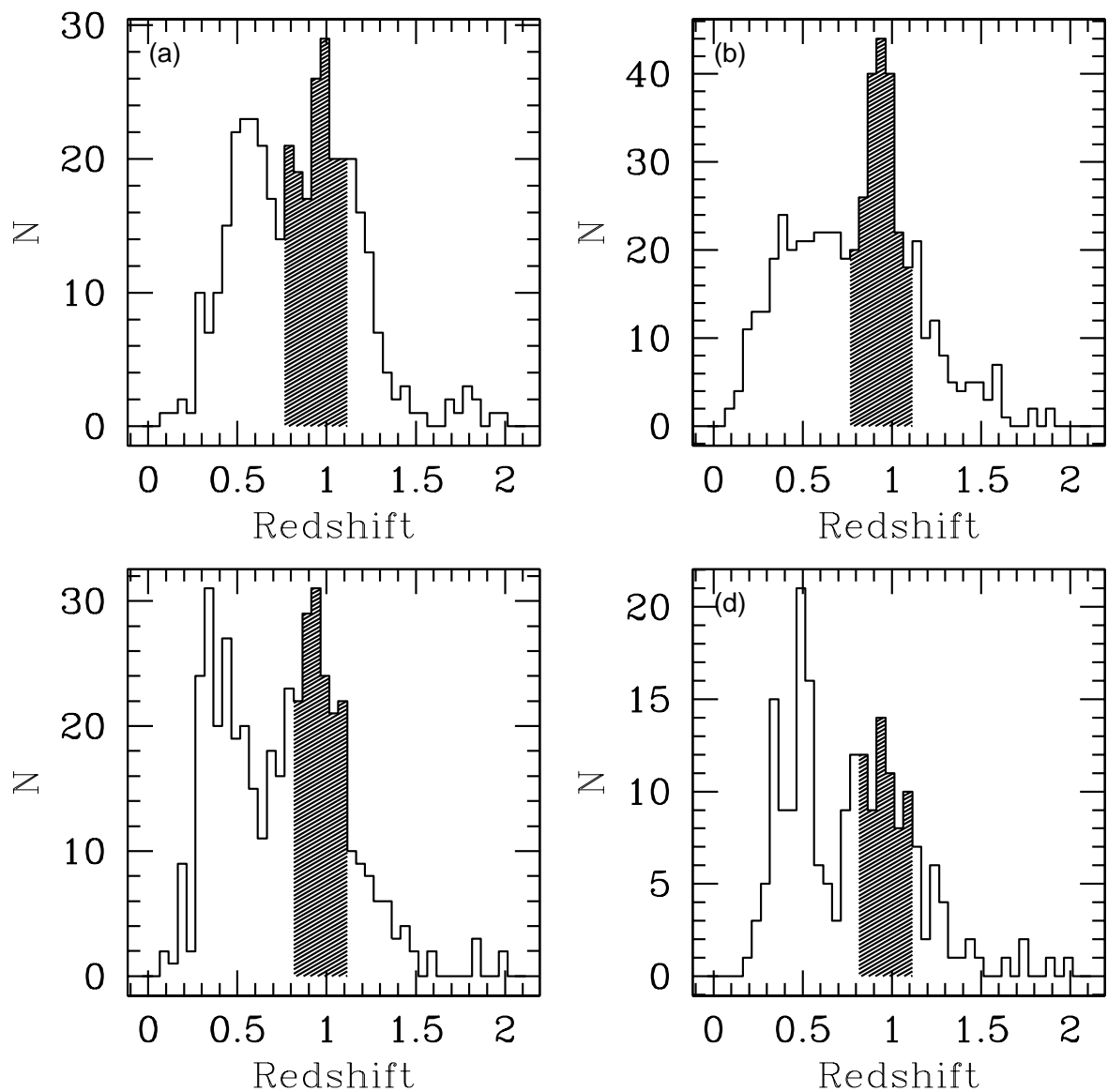} \caption{\label{dr}{Distribution of
photometric redshifts for each field containing a quasar pair:
a) \qpa, b) \qpb, c) \qpc, and d) \qpd.
The hatched area in each panel corresponds to $z_{pair} \pm \sigma_z$.}}
\end{figure}

We present  in Figure \ref{dr} the  galaxy redshift distribution in the field
of each quasar pair. All fields show a peak in the interval
$z \in[z_{pair}-\sigma_z,z_{pair}+\sigma_z]$.
We now analyse some properties of the galaxy
distribution in this redshift interval aiming at constraining the nature of the
environment inhabited by the quasar pairs of our sample.

\subsection{Galaxy overdensities}

We must know the expected number of galaxies in this interval,
to verify the significance of the galaxy excess around $z_{pair}$.
For this estimate we have assumed that the HHDFN region is representative of
the overall galaxy distribution. This sample is appropriate for this
analysis because its photometric depth is comparable to that of our fields.
Photometric redshifts were obtained with  the method discussed in the
previous section. We then defined the galaxy overdensity in the
interval $z_{pair} \pm \sigma_z$ as
\begin{equation}
\label{ec}
\delta = \frac{n_{pair}-n_{H}}{n_{H}}
\end{equation}
where $n_{pair}$ and $n_{H}$
are the number densities of galaxies in this redshift interval for a given
field and for the HHDFN, respectively.
We have considered as galaxies in HHDFN 
all objects with $z_{phot} > 0$.

Values of $\delta$ for each pair are shown in Table \ref{prop}.
Errors in $\delta$ were determined assuming Poissonian errors for $N_{pair}$
and $N_H$. The overdensity $\delta$ ranges from 0.6 to 1.6 and is
significant in all cases.
Note that these results are affected by cosmic variance, since
we have used only one reference field and the  area occupied by HHDFN
(0.2 square degrees) is very small, so that $n_H$ is
affected by the galaxy clustering in the HHDFN region.
We have arrived at similar results using the VIMOS VLT Deep Survey
around the Chandra Deep Field South \citep{vimos} and 
the Gemini Deep Deep Survey \citep{gdds} regions.

\subsection{Distribution of galaxies}

In order to investigate the clustering properties of the galaxies in
the chosen redshift interval around a quasar pair, we calculated the
median projected distance between galaxies and compared them
with the same quantities obtained with 1000 simulations of
random uniform galaxy distributions with the same
number of objects and the same projected area of the observed fields. We
may then define a confidence level, $CL$, that a field presents a
galaxy distribution more clustered than an uniform distribution:

\begin{equation}
\label{clex}
CL = \frac{N (\Delta \theta > \Delta \theta_{f})}{N_s}
\end{equation}
where $N (\Delta \theta > \Delta \theta_{f})$ is the number of simulated
fields with median projected distances larger than that of the
observed fields and $N_s$ is the total number of simulations.

We summarize the results of this analysis in Table \ref{prop}.
Two pairs are strongly clustered (\qpb, \qpc), one is moderately clustered
(\qpa) and one (\qpd) is not clustered at all.

\subsection{Richness}

In order to estimate the richness of our fields, we have adopted an approach
similar to the traditional Abell's richness criterion \citep{abell58}, defined
as the number of galaxies brighter than $m_3 + 2$ (where $m_3$ is the
magnitude of the third brightest cluster member) within a radius of
$1.5 ~h_{100}^{-1}$ Mpc of the cluster center. A cluster is considered
rich if it contains more than 30 galaxies according to such a definition.

The Abell's radius considering the cosmological model adopted here 
is 2.1 Mpc. Assuming
that the brightest galaxy in the redshift interval $z_{pair} \pm
\sigma_z$ is the brightest cluster galaxy, we computed the number of
galaxies brighter than $i'_3+2$ by scaling their number in each
field to the Abell area ($N^{esc}=N/\Sigma$, 
where $\Sigma=A_{par}/A_{Abell} \sim 0.5$). 
This result was corrected for contamination
due to background/foreground objects using counts in the HHDFN
region. The results are shown in Table \ref{prop}. All but one of
the putative clusters are rich, according with this criterion.  The
field of \qpa ~seems to be the poorest of our four fields, and is
poor also with Abell's criterion. However, this is the pair with the
brightest galaxy in the corresponding redshift interval among all
quasar pairs, and its poorness may be an effect of galaxy counts,
since they grow strongly for increasing magnitude. 
Furthermore, note that our fields are smaller
than the Abell's radius, then the quasars could be in a poor cluster,
group, or in the neighborhood of a cluster.
On the other
hand, the pair \qpd, which is not rich by the results of Sections 5.1 and 5.2, has 
 R=0 in Abell's classification. It is, then, appropriate to look
for other richness estimators to confirm or not these results.

Another useful richness indicator is the number of bright galaxies,
assumed here as those brighter than $i'^*+1$ present in the field. 
This number is estimated
for galaxies in the pair redshift interval and is corrected with the
corresponding HHDFN counts (scaled to the field area). The results are
also presented in Table \ref{prop}. All fields seems to contain a
considerable number of bright galaxies.

It is interesting to compare our results with those obtained by
\citet{pos02}. These authors studied a variety of Abell-like
richness indicators in an I-band cluster survey. One of these
indicators, $N_{A,0.5}$, is defined as the number of galaxies with
magnitude between $m_{3}$ and $m_{3}+2$ within a radius of 666
$h_{75}^{-1}$ $kpc$. They show
that this indicator is related to Abell's richness, $N_A$, as
$N_{A,0.5} \sim 0.44 N_{A}$. We have used this relation to estimate
Abell's richness from $N_{A,0.5}$. For 31 clusters with redshifts
between 0.9 and 1.0, we obtain $N_{A} = 54$ galaxies.
That means that  fields have a richness $R \sim 1$, which may be
compared with the numbers present in Table \ref{prop}: only  \qpa
~seems poorer than the clusters at comparable redshift studied by
\citet{pos02}.

\subsection{The red sequence}

The red sequence is a characteristic of the color-magnitude
diagrams of early-type galaxies of groups
and clusters. In a color-magnitude diagram these galaxies have very
similar colors following a linear relation and their integrated colors are
progressively bluer for weaker magnitudes. This relation is also known as
color-magnitude relation (CMR).

\begin{figure}[h!]
\plotone{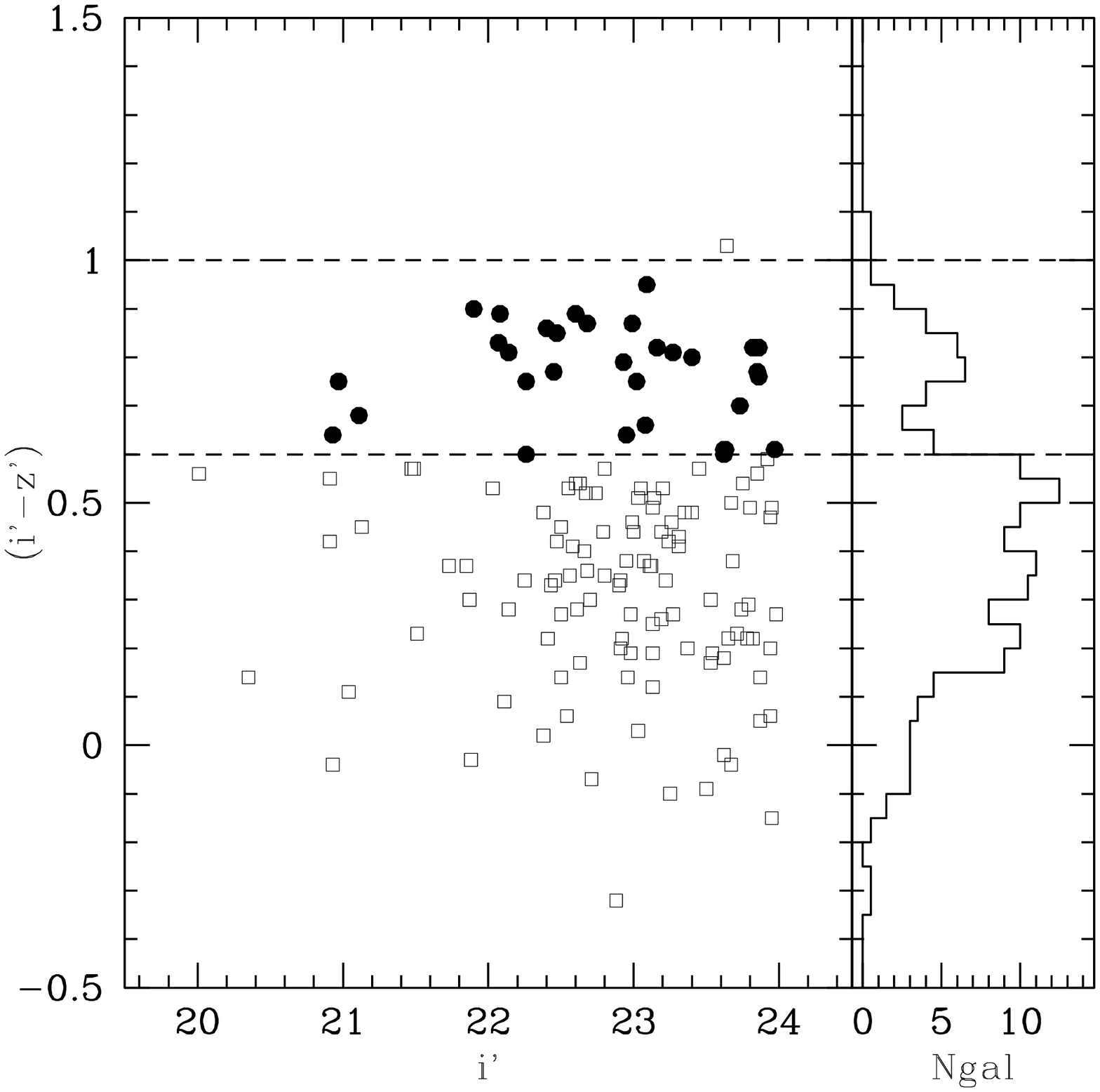}\\
\plotone{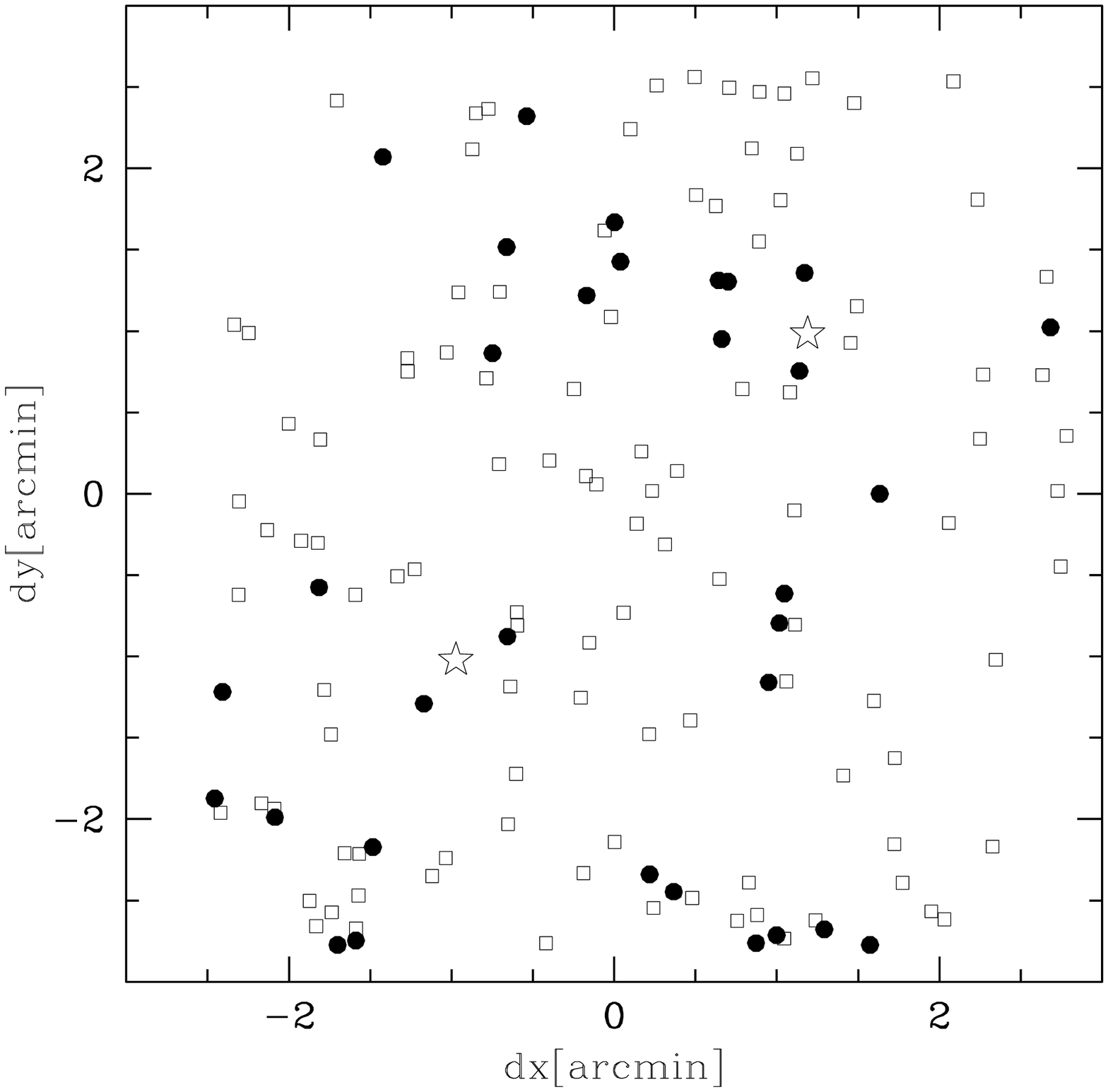}

\caption{\label{cla}{Results for the pair \qpa. 
Galaxies within $z_{pair} \pm \sigma_z$. Top-left:
color-magnitude diagram; galaxies in the red sequence are represented as
filled circles. Top-right: histogram of the color distribution.
Bottom: projected distribution of galaxies. The
quasar pair is represented by stars and the galaxies in the red sequence
as filled circles.}}
\end{figure}

\begin{figure}[h!]
\plotone{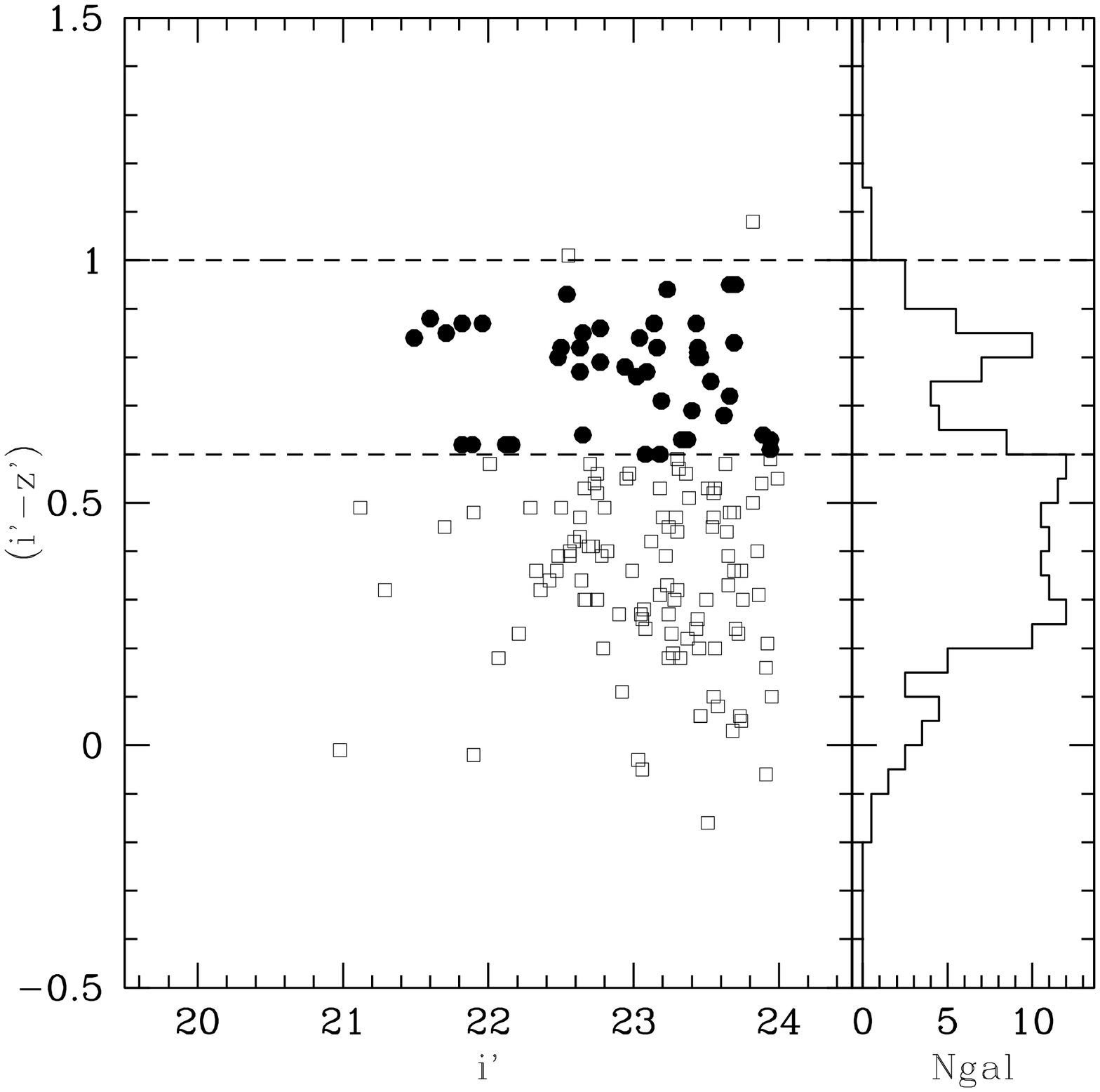}\\
\plotone{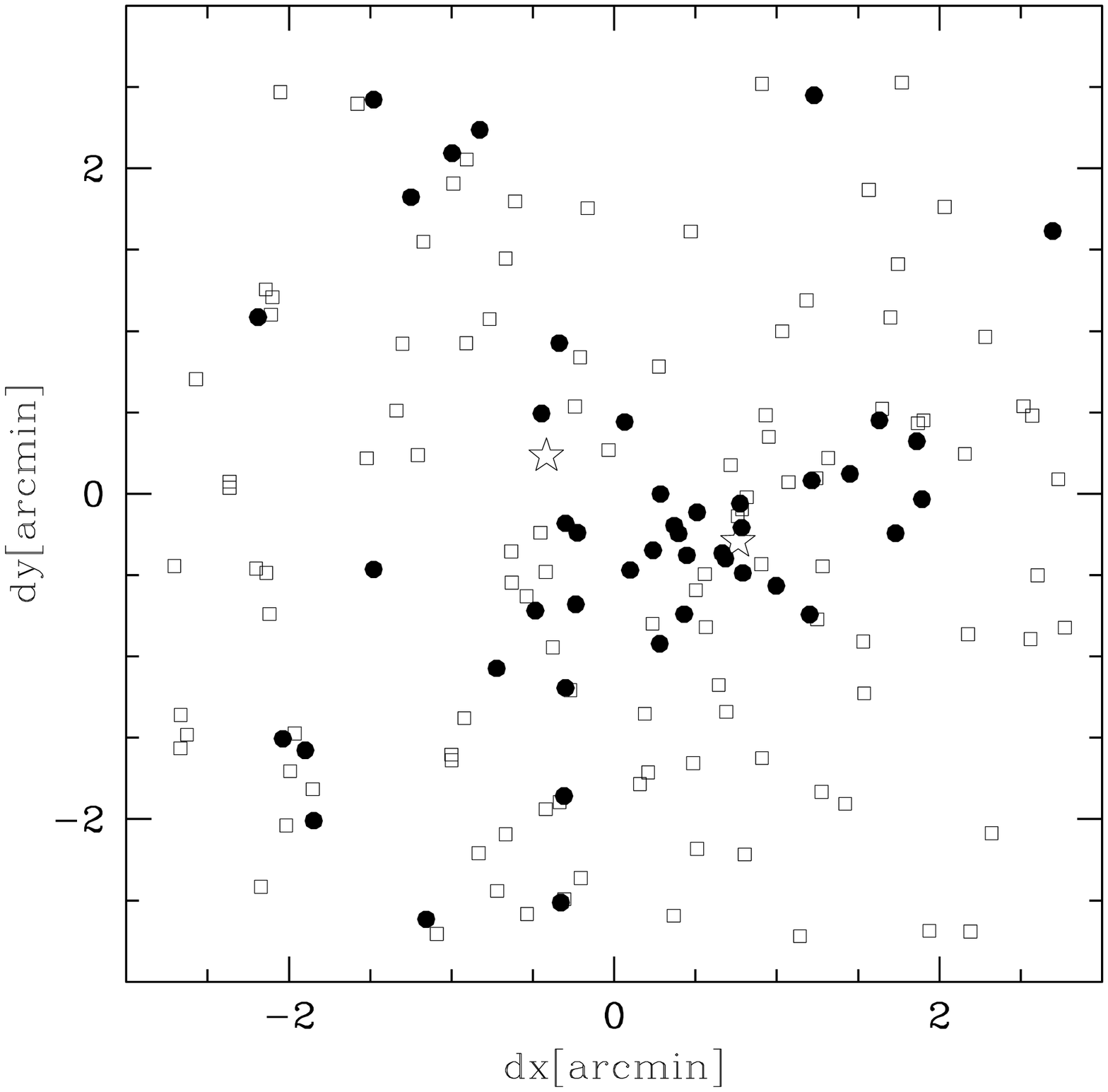}

\caption{\label{clc}{Same as Fig. \ref{cla} but for \qpc.}}
\end{figure}

We have examined the red sequence in the $(i'-z') \times i'$
diagram of galaxies in the redshift interval of each pair. 
The use of the color $(i'-z')$ is based on its capability to 
identify early-type galaxies, since at $z \sim 1$ the 
4000 \AA~ break lies in the $i'$ band and consequently 
the early-type galaxy color $(i'-z')$ are very red.
The fields around
quasar pairs \qpa ~and \qpc ~(Figures \ref{cla} and \ref{clc} -
top-right) present a peak in the color distribution at $0.6 \le
i'-z' \le 1.0$. 
Comparing our data with a
similar distribution for HHDFN, we verify that these peaks represent
an excess of 1.7 $\sigma$ and 3.3 $\sigma$, respectively. 
Therefore, in this interval, we would expect to find a red sequence
in the color-magnitude diagram. Indeed, for \qpc ~ we note
clearly that the galaxies form a red sequence in $i'-z' \sim 0.8$
(Figure \ref{clc} - top-left), the value obtained by
\citet{tanaka06} in spectroscopically confirmed structures at $z \sim
0.9$. Besides, if we consider the projected distribution of these
red-sequence galaxies (Figure \ref{clc} - bottom),  we notice
that they have a filamentary-like distribution similar to what is
observed in other $z \sim 1$ clusters, and considered typical of
clusters in process of formation (e.g., Toft et al. 2003).

\begin{figure}[h!]
\plotone{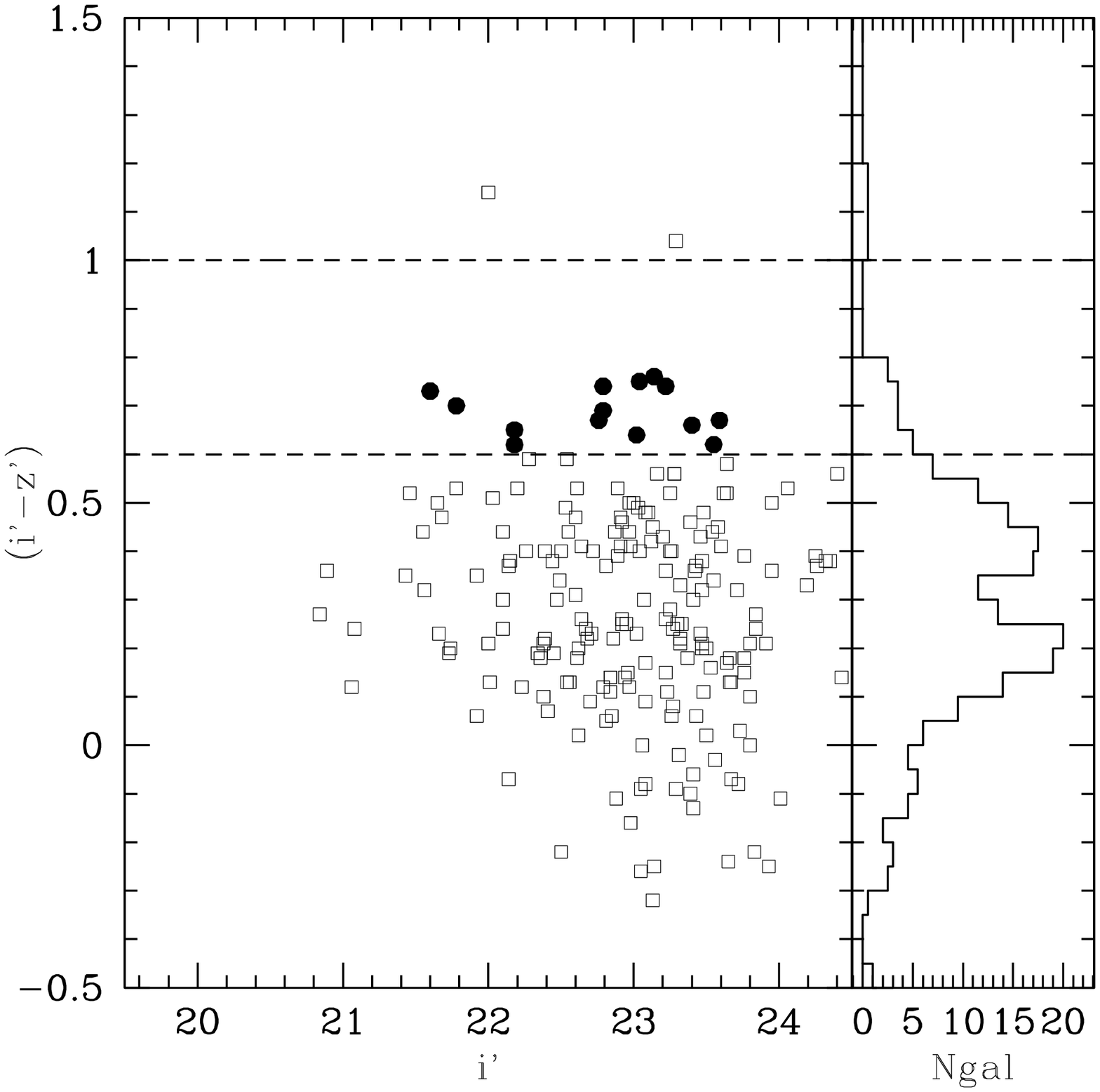}\\
\plotone{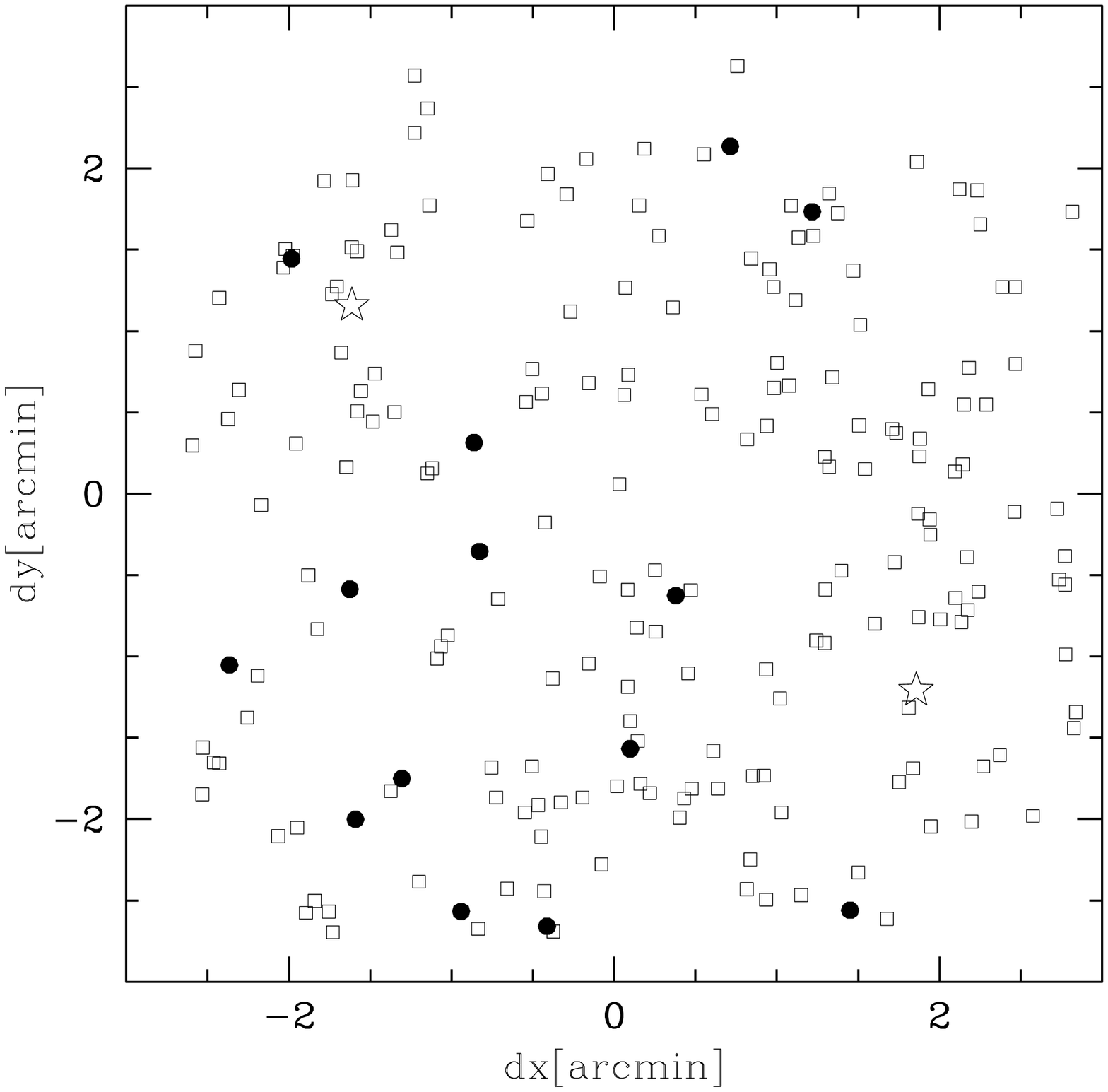}

\caption{\label{clb}{Same as Fig. \ref{cla} but for \qpb.}}
\end{figure}

\begin{figure}[h!]
\plotone{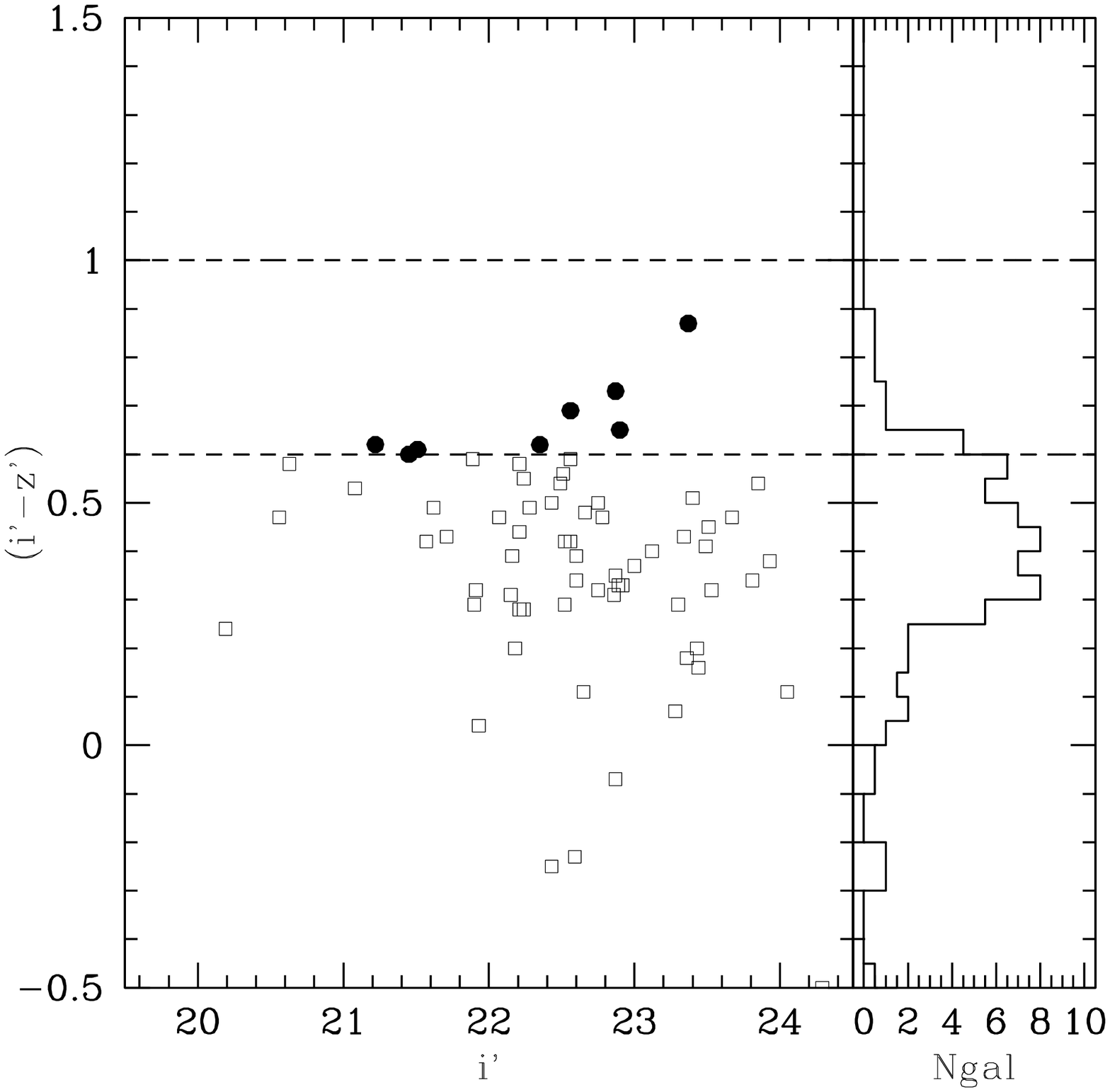}\\
\plotone{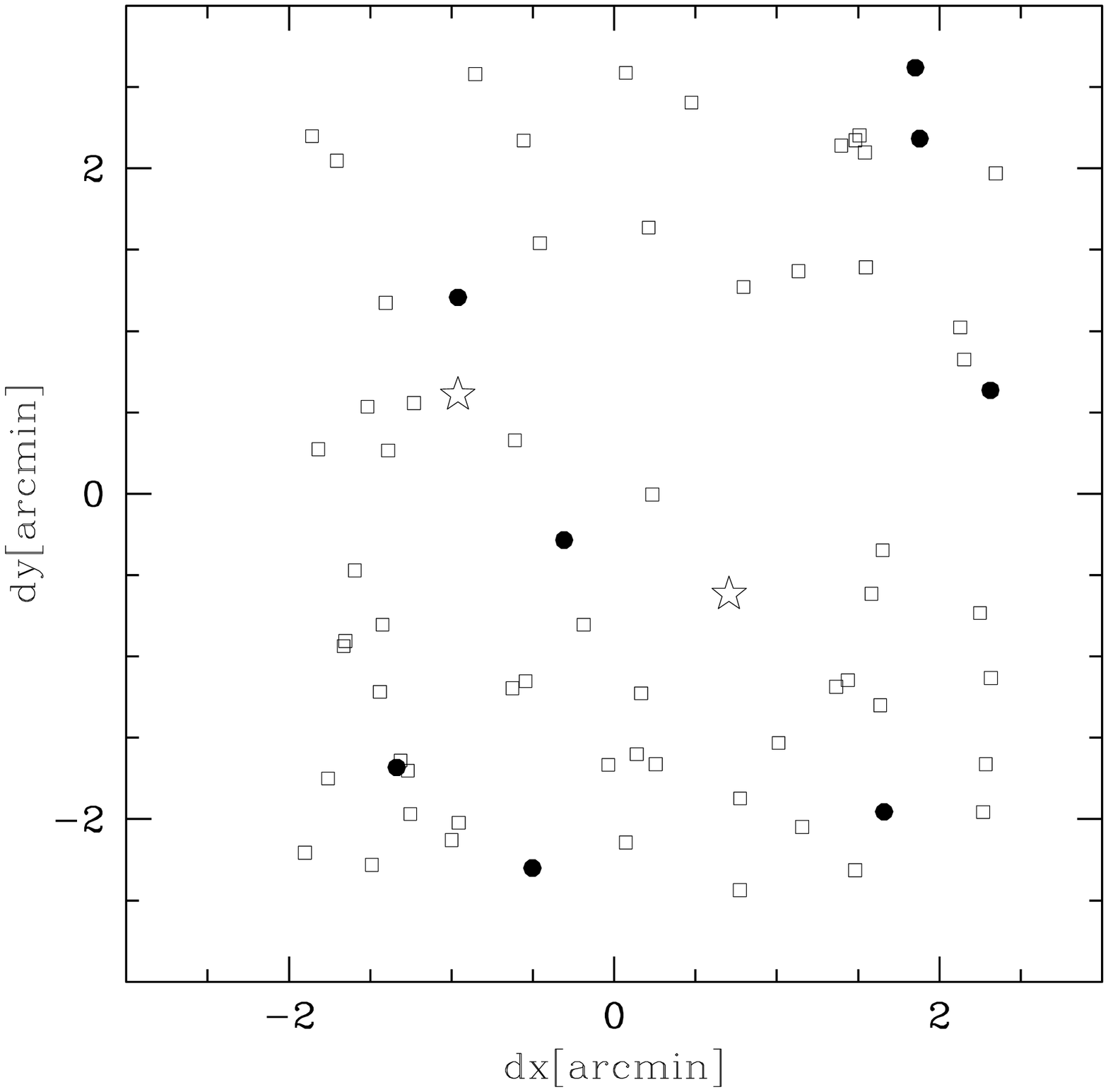}

\caption{\label{cld}{Same as Fig. \ref{cla} but for \qpd.}}
\end{figure}

The red galaxies of \qpa~ present a broad distribution in the
color-magnitude diagram (Figure \ref{cla} - top-left).
They also present a clump-like projected distribution
(Figure \ref{cla} - bottom). The other two fields have less-significant
red sequences (Figures \ref{clb} and \ref{cld}). The cluster CL1604+4321,
at $z \sim 0.9$, the less massive of the clusters studied by
\citet{home06}, presents a lack of bright elliptical galaxies
($\sim M^*$). The authors suggest that this cluster
has not yet had time to complete the red sequence. This may be also the
case for the structures associated to \qpb ~and \qpd.

\subsection{Properties of the fields around quasar pairs}

The properties of the environment associated with each quasar pair are
summarized in Table \ref{resumo}. We now discuss each pair individually.

\subsubsection{\qpa}

This quasar pair is formed by \qaa ~(a radio-quiet object) and \qab
~(a radio-loud quasar) at redshifts 0.925 and 0.933, respectively.
They have an angular separation of 177 arcsec, corresponding to 1.4
Mpc in the adopted cosmology. Its density contrast is the smallest
among all quasar pairs. However, the galaxy distribution 
analysis shows that the galaxies in this field 
are clustered at some degree, i.e., 
the median projected distance between galaxies is smaller than in 
a random uniform field in 67 \% of the simulations. 
This field has been classified as poor with
the Abell's criterion, and we have found 20 galaxies with magnitude
$i'<i'^*+1$. The galaxy color distribution shows a prominent peak in
$ 0.6 \le i'-z' \le 1.0$, corresponding to the red sequence. The red
galaxies present a clumpy distribution, but without central
condensation. The presence of a significant amount of early-type galaxies plus the
relative poorness of this field indicate that this can be the seed
of a structure that can became a rich galaxy cluster at $z=0$.

\subsubsection{\qpb}

\qba ~and \qbb ~constitute this pair; the first is radio-loud and
the second is radio-quiet. The projected separation between them is
252 arcsec, or 2.0 Mpc, with redshifts 0.932 and 0.934,
respectively. The redshift interval $z_{pair} \pm \sigma_z$ shows
the largest galaxy excess among all quasar pairs discussed in this
work. 
The median projected distance between galaxies resulted smaller than in
a random uniform field in 98.5 \% of the cases, meaning that these galaxias are
strongly clustered. This is the richest field in our sample
accordingly to Abell's criterion and also the one with the largest
number of bright galaxies ($i'<i'^*+1$), however its red sequence 
is modest and the red
galaxies do not present a clustered distribution. Its richness and
number of bright galaxies are the major indications that this quasar
pair is probably in a galaxy cluster.

\subsubsection{\qpc}

This quasar pair is formed by a radio-loud quasar (\qca)
and a radio-quiet quasar (\qcb) at redshifts 0.958 and 0.956, respectively.
The angular separation of 77 arcsec (0.6 Mpc) is the smallest  of the sample.
The overdensity in the redshift interval is significant as for the
other pairs. The galaxy
distribution is the most clustered of all samples, accordingly to the
CL values in Table  \ref{prop}. The field is rich by Abell's
criterion, but presents only 12 bright galaxies. The red sequence is
clearly present in the color-magnitude diagram at $i'-z' \sim 0.8$.
The red galaxies present a filamentary-like distribution and there is
a galaxy excess around the radio-loud quasar. These results indicate
that \qpc ~is indeed a rich cluster.

Moreover, we have verified that  \qpc ~has been serendipitously
detected (but unreported) in X-ray with a pointed \textit{ROSAT}
PSPC observation of 6.6~ks. We have estimated the  bolometric X-ray
luminosity assuming that all detected flux (background corrected)
comes from the ICM: $L_{X, \rm bol} \sim 5 \times
10^{45}\,$ergs~s$^{-1}$. Such luminosity is well above a typical
cluster X-ray luminosity and may be contaminated by the X-ray
emission from one or both quasars. On the other hand, the typical quasar
X-ray luminosity [2--10 keV] is around $10^{44}\,$erg~s$^{-1}$, thus
the quasars in the pair may not account for all X-ray emission.
Besides, the total emission [0.5--8.0 keV] within 3 arcmin is about
10 times higher than the typical quasar emission in this band;
therefore the observed X-ray flux is consistent with  emission from
quasars and a possible cluster around them. This is the single pair
in our sample detected in X-rays so far.
The other fields were not detected in the
Rosat All Sky Survey, nor in the pointed observations.

\subsubsection{\qpd}

This pair is formed by radio-quiet quasars. \qda ~has $z = 0.974$
and \qdb ~has $z = 0.968$. The separation between them is 144 arcsec
(1.1 Mpc). This field shows a significant overdensity, no
clustering, and no red sequence. It also seems rich with Abell's
criterion and has a comparatively large number of bright galaxies.
The evidence for the presence of a rich cluster at the redshift of
the quasar pair is not as compelling in this case, compared with the
other 3 pairs.

\section{Summary}

We have studied the environment traced by quasar pairs at $z \sim 1$, using
images in $g'$, $r'$, $i'$, and $z'$ bands obtained with GMOS at Gemini
North and South. In order to identify galaxies in a redshift interval close
to that of the quasar pairs, we have estimated photometric redshifts with
the LWR method, using ACS-GOODS data as a training set.
The rms dispersion of the difference between our photometric redshift and the spectroscopic
redshift in the training set is $\sigma_z = 0.16$. We have adopted the
interval $z = z_{pair}$ $\pm$ $\sigma_z$ for the analysis of the
pair environment.

When compared with the HHDFN region, all fields show a significant
overdensity in the redshift interval of the pair. In all cases
this excess is larger than 3.5 $\sigma$.

We investigated the clustering of the galaxies near the pair by estimating
a confidence level, $CL$, that the galaxies are more concentrated than in a
uniform distribution. We have also estimated the richness of each
redshift interval with a variant of Abell's criterion, as well as by the
number of bright galaxies. We verified whether a red sequence is
present and the form of the projected distribution of red galaxies.

The analysis indicates that probably three out of our four quasar pairs 
are members of galaxy custers. For one of the pairs we did not find
strong evidence for it: \qpd~ could be in a poor cluster, group, or in the neighborhood
of a cluster, since our fields are lesser than Abell's radius.
Taken at face value, this result shows that
quasar pairs are indeed good tracers of the large scale structure at
high $z$. However, with only four quasar pairs in our 
sample we are not able to say at what level targeting a quasar pair increases 
the probability of finding a rich galaxy cluster as compared to targeting a 
single quasar. A study of larger and homogeneous samples would be necessary 
to clarify this point. An extension of our work to other redshifts may be 
also useful and may provide interesting clues on the evolution of large-scale 
structure and galaxy clustering.

\acknowledgments
This work is based on observations obtained at the Gemini Observatory, which is operated by the
Association of Universities for Research in Astronomy, Inc., under a cooperative agreement
with the NSF on behalf of the Gemini partnership: the National Science Foundation (United
States), the Particle Physics and Astronomy Research Council (United Kingdom), the
National Research Council (Canada), CONICYT (Chile), the Australian Research Council
(Australia), CNPq (Brazil) and CONICET (Argentina).

We would like to thank the Gemini staff for obtaining the observations
in queue mode. We also thank G. B. Lima Neto for many useful discussions,
and the anonymous referee, whose comments helped to
improve this paper.
We are grateful for the support provided by the Brazilian agencies CNPq and
FAPESP.
We made use of the NASA/IPAC Extragalactic Database, which is
operated by the Jet Propulsion Laboratory, California Institute
of Technology, under contract with NASA.

\clearpage
%\clearpage

\clearpage
%\clearpage

\clearpage
%\clearpage

\clearpage
\epsscale{.6}

%\clearpage

%\clearpage

%\clearpage

\clearpage
\begin{deluxetable}{cccccc}
\tabletypesize{\scriptsize} \tablecaption{\label{car} Sample
Characteristics} \tablewidth{0pt} \tablehead{ \colhead{Quasar Names}
& \colhead{$\alpha$} & \colhead{$\delta$} & \colhead{z} &
\colhead{$\Delta \theta$} & \colhead{Quasar Pair Name} \\
  & (J2000) & (J2000) &  & ($arcsec$) &  
}

\startdata

J131046+0006\tablenotemark{*} & 13 10 46.2 &  00 06 33 & 0.925 & 177  & \qpa \\
J131055+0008  & 13 10 55.9 &  00 08 14 & 0.933 &      &        \\
           &        &       &   &      &        \\
J135457-0034  & 13 54 57.2 & -00 34 06 & 0.932 & 252   & \qpb \\
J135504-0030\tablenotemark{*} & 13 55 04.7 & -00 30 20 & 0.934 &      &      \\
           &        &       &   &      &        \\
Q 0107-0235   & 01 10 13.2 & -02 19 53 & 0.958 & 77    & \qpc \\
PB 6291\tablenotemark{*}      & 01 10 16.3 & -02 18 51 & 0.956 &      &      \\
           &        &       &   &      &        \\
J011441-3139\tablenotemark{*} & 01 14 41.8 & -31 39 25 & 0.974 & 144 & \qpd \\
J011446-3141\tablenotemark{*} & 01 14 46.4 & -31 41 31 & 0.968 &      &      \\
\enddata

\tablenotetext{*}{Radio-quiet quasars}

\end{deluxetable}
%\clearpage

\clearpage
\begin{deluxetable}{lcccccc}
\tabletypesize{\scriptsize} \tablecaption{\label{obs2} Observations}
\tablewidth{0pt} \tablehead{ \colhead{Pair} & \colhead{Telescope} &
\multicolumn{4}{c}{$t_{exp}$ (seconds)}
&  \colhead{Identification Number} \\
& & \colhead{$g'$} & \colhead{$r'$} &  \colhead{$i'$} &
\colhead{$z'$} & \colhead{}}

\startdata

\qpa & Gemini N &  9 $\times$ 300.0 & 6 $\times$ 200.0 & 11 $\times$ 350.0 & 8 $\times$ 450.0 & GN-2003A-Q-2/GN-2005A-Q-19 \\
\qpb & Gemini N & 13 $\times$ 300.0 & 6 $\times$ 200.0 &  6 $\times$ 350.0 & 7 $\times$ 450.0 & GN-2003A-Q-2/GN-2005A-Q-19 \\
\qpc & Gemini N & 10 $\times$ 300.0 & 6 $\times$ 200.0 &  8 $\times$ 350.0 & 8 $\times$ 410.0 & GN-2003B-Q-2/GN-2004B-Q-24 \\
\qpd & Gemini S &  7 $\times$ 300.5 & 6 $\times$ 200.5 &  7 $\times$ 350.5 & 7 $\times$ 410.5 & GS-2003B-Q-3/GS-2004B-Q-17 \\

\enddata

\end{deluxetable}
%\clearpage

\clearpage
\begin{deluxetable}{cccccccccc}
\tabletypesize{\scriptsize} \tablecaption{\label{prop} Clustering
Properties} \tablewidth{0pt} \tablehead{ \colhead{Pair} &
\colhead{$\delta$} & \colhead{$\Delta \theta_{median}$} &
\colhead{$CL_{median}$} & \colhead{$i'_3$} &
\colhead{$N(i'<i'_{3}+2)$} &
\colhead{$N^{esc}(i'<i'_{3}+2)$} & $N(i'<i'^*+1)$\\
 &  & (arcmin) & (\%) & & & &
}

\startdata

\qpa & 0.58 $\pm$ 0.14 & 2.7 &  67.0  & 20.35  &   6 ($R<0$)  &  13 ($R<0$) &  20 \\
\qpb & 1.59 $\pm$ 0.19 & 2.6 &  98.5  & 21.06  &  95 ($R=2$)  & 203 ($R=4$) &  58 \\
\qpc & 0.70 $\pm$ 0.14 & 2.4 & 100.0  & 21.29  &  35 ($R=0$)  &  72 ($R=1$) &  12 \\
\qpd & 0.86 $\pm$ 0.23 & 2.8 &   0.5  & 20.63  &  34 ($R=0$)  &  95 ($R=2$) &  36 \\

\enddata

\end{deluxetable}
%\clearpage

\clearpage
\begin{deluxetable}{ccccccc}
\tabletypesize{\scriptsize} \tablecaption{\label{resumo} Summary of
the quasar pair properties} \tablewidth{0pt} \tablehead{
\colhead{Pair} & \colhead{$\delta$} & \colhead{$CL$} &
\colhead{$N_A$} & \colhead{RCM} &  \colhead{C/F\tablenotemark{a}} &
X-rays }
\startdata
\qpa & ok & ok &  x &  ok & ok  &  -- \\
\qpb & ok & ok & ok &   x &  x  &  -- \\
\qpc & ok & ok & ok &  ok & ok  & ok \\
\qpd & ok &  x & ok &   x &  x  &  -- \\
\enddata
\tablenotetext{a}{Cluster-like or filament-like distribution.}
\end{deluxetable}
%\clearpage

\end{document}